\documentclass[aps,prl,showpacs,twocolumn,groupedaddress,amsmath]{revtex4}
\usepackage{graphicx}
\usepackage{dcolumn}
\usepackage{bm}
\usepackage{epsfig}
\begin{document}
        
\title{Scaling in the Rubinstein--Duke Model for Reptation}

\author{Andrzej Drzewi\'nski$ ^{1,2}$ and J.M.J. van Leeuwen$ ^3$}
\affiliation{$ ^{1}$Czestochowa University of Technology, 
Institute of Mathematics and Computer Science,
ul.Dabrowskiego 73, 42-200 Czestochowa, Poland}
\affiliation{$ ^{2}$Institute of Low Temperature and Structure Research, 
Polish Academy of Sciences,
P.O.Box 1410, 50-950 Wroc\l aw 2, Poland}
\affiliation{$ ^3$Instituut-Lorentz, University of Leiden, P.O.Box 9506, 
2300 RA Leiden, the Netherlands}

\date{\today} 
  
\begin{abstract}
We consider an arbitrarily charged polymer driven by a weak field  through a gel 
according to the rules of the Rubinstein--Duke model. The probability distribution in the
stationary state is related to that of the model in which only the head is charged. 
Thereby drift velocity, diffusion constant and orientation of any charged polymers
are expressed in terms of those of the central model. Mapping the problem on a random walk
of a tagged particle along a one-dimenional chain,
leads to a unified scaling expression for the local orientation. It provides also an elucidation
of the role of corrections to scaling.
\end{abstract}


\maketitle
The basic ingredient of the physics of polymers is scaling,
i.e. the analysis of the properties as function of the length $N$ of the polymer chain 
\cite{deGen}. While most of the leading scaling behavior is understood, the comparison
between theory and experiment is hampered by large corrections to scaling. 
A typical example is the renewal time $\tau$ for polymers dissolved in a gel.
Theory predicts behavior $\tau \sim N^3$, 
while experiments seem to converge on $\tau \sim N^{3.4}$ \cite{deGen}. The discrepancy
has been blamed on large corrections to scaling, blurring the true asymptotic behavior
\cite{Doi,Rubinstein}.
This suggestion has been given a firm basis by the analysis of Carlon et. al. \cite{Carlon} 
of the renewal time in the Rubinstein-Duke (RD) model for reptation. Polymers up to thousands
of base pairs may still have large corrections to the asymptotic scaling behavior.
Since they find that deviations from the leading scaling behavior decay as $1/\sqrt{N}$,
their work indicates that the correction to scaling exponent is -1/2.

Sofar the scaling analysis has been restricted to global properties, such as renewal time and
diffusion coefficient. Equally interesting and more informative, are scaling
properties related to the position in the chain. As example consider a neutral polymer 
dissolved in a gel with a magnetic beat attached to the head of the polymer. The polymer
can be pulled through the gel by a magnetic field. It is the analogon of the more common
electrophoresis (EP) and has been named magnetophoresis (MP) \cite{Schutz}.
The field orients the polymer links, at the head stronger than at the tail. 
On top of this overal effect is a subtile scaling behavior near head and tail, as we will 
show. Similar effects occur in electrophoresis, which is a central
tool of DNA fingerprinting. DNA, being an acid, acquires as local charge, when dissolved
in a gel like agarose. These local charges are evenly spread over the chain and an electric
field pulls equally strong at all elements of the polymer. The local orientation of this
chain has an even more interesting scaling behavior.

The motion of a polymer is far too complicated to be taken into full detailed account.
Therefore lattice models have been designed for reptation, in which 
the polymer is viewed as a chain of hopping reptons \cite{deGen2},
confined to a tube of pores in the gel.
Among them the RD model stands out by the  simplicity of the motion rules
{\cite{Rubinstein,Duke,Widom}.
Although the model is a rather crude simplication of the reality, it captures
a number of basic features of reptation \cite{Barkema}.
This paper is concerned with scaling effects involving the position along the chain, as they
emerge in the RD model. 

An attractive aspect of the RD model is the fact that its hopping operator 
is a  one-dimensional spin operator, albeit a non-hermitian hamiltonian.
This allows to apply the Density Matrix Renormalization Group (DMRG) technique 
to study the stationary state of the Master Equation \cite{White,Carlon}. 
The DMRG method gives precision data for the whole scala of chains up to lengths of 
150 reptons, which make it ideally suited for finite size analysis. 
This has an advantage over direct simulations of the model, which are slow, due to the long 
renewal time, and limited by statistical errors.

In this paper we linearize the Master Equation for the probability distribution with respect
to the driving field and study the solution for an arbitrarily charged polymer. First we
relate the probability distribution of the general case
to that for the MP model. This enables us to link the
properties of the original RD model, with equal charges on all reptons, to those
of the MP model. We map the MP model on the problem of a random walker on a
one-dimensional chain. Going back and forth from the EP  and 
MP variant, we derive scaling properties for the various regions of the local orientation.

The RD model views the polymer as a a one-dimensional string of reptons connected
by links. The reptons are located in the cells of a $d$-dimensional
hypercubic lattice with the field along the body diagonal \cite{Widom}. 
We number the reptons from 1 (tail) to $N$ (head).
The reptons hop independently from each other, under the constraint that the links always 
connect two reptons in adjacent cells, or two reptons in  the same cell. 
Only the projection of a link on the field direction is important for the
probability distribution. A configuration of the polymer can be 
represented by a vector ${\bf y} = (y_1, \cdots, y_{N-1})$, 
of link variables, with $y_j = 0,\pm 1$, measuring the distance between
the reptons $j$ and $j+1$ along the field direction \cite{Widom}.

The Master Equation for the  stationary probability distribution $P({\bf y})$ has the form 
\begin{equation} \label{a1}
\sum_{\bf y'} \left[ W({\bf y} | {\bf y}') P ({\bf y}') - W({\bf y}' | {\bf y}) P ({\bf y}) \right]
\equiv {\cal M} P ({\bf y}) = 0.
\end{equation} 
$W({\bf y} | {\bf y}')$ is the transition rate from configuration ${\bf y}'$ to ${\bf y}$. 
A move of repton $i$ in the direction of the field has a bias $B_i  = \exp(q_i \epsilon /2)$,
where $\epsilon$ measures the strength of the field and $q_i$ the charge of the repton. 
Reptons moving opposite to the field are biased with $B_i ^{-1}$. Although the influence of the
embedding dimension has interesting aspects \cite{Carlon2}, we confine ourselves in this paper 
to $d=1$, for which the most extensive DMRG results are obtained, 
yielding the most accurate scaling data.

We consider the field strength $\epsilon$ as a small parameter ($\epsilon N < 1$); 
it is experimentally the most relevant regime. 
For $\epsilon = 0$ the stochastic matrix is symmetric and has the solution 
$P^0 ({\bf y}) = 3^{-(N-1)}$, all $3^{N-1}$ configurations being equally probable. 
Expanding the Master Equation in powers of $\epsilon$
\begin{equation} \label{a2}
{\cal M} = {\cal M}^0 + \epsilon {\cal M}^1 + \cdots, \quad  P({\bf y} ) = 
P^0 ({\bf y}) + \epsilon P^1 ({\bf y}) + \cdots, 
\end{equation}  
leads to the lowest order  equation ${\cal M}^0 P^0 = 0$ and to the first order equation
\begin{equation} \label{a3}
{\cal M}^0 P^1  + {\cal M}^1 P^0 = 0.
\end{equation} 
This paper is concerned with its solution.

The remarkable point is that the dependence on the charge distribution can be made explicit. 
\begin{equation} \label{b1}
P^1 ({\bf y}) =- \sum^{N-1}_{j=1} \left(\sum^j_{i=1}  q_i \right)  y_j P^0 ({\bf y}) + 
\left(\sum^N_{i=1} q_j \right) P^1_{MP} ({\bf y}).
\end{equation} 
Here $P^1_{MP} ({\bf y})$ is the first order distribution of the MP model with a unit
charge on the last repton only. The proof follows by substitution in (\ref{a3}). 
So all cases are reduced to this central model.

A direct consequence of this relation is that the drift velocity
$v$ of an arbitrary $q_i$ is related to that of the MP model as
\begin{equation} \label{b2}
v = \left(\sum^N_{i=1} q_j \right) v_{MP}
\end{equation} 
Thus the drift velocity depends only on the total charge.
This was anticipated in \cite{Carlon3}. E.g. the drift velocity of the EP
model is $N$ times that of the MP model.

A nice numerical illustration of relation (\ref{b1}) is given by the calculation of the local
orientation $\langle y_j \rangle$. This is a first order effect and (\ref{b1}) gives a
relation e.g. between the EP model (all $q_i=1$) and the MP model (only $q_N = 1$). 
Both models have been treated with DMRG \cite{Carlon,Carlon2,Carlon3}. 
In Fig. \ref{compar} we show the values 
for the MP model and the equivalent combination for the EP model. 
The correspondence is perfect. It shows that for both calculations $\epsilon$
is small enough to guarantee that the weak field limit applies. It also is a proof of the
accuracy of the DMRG method, since the calculations were performed prior to the derivation
of relation (\ref{b1}).
\begin{figure}[h]
 \centering
 \includegraphics[width=7cm]{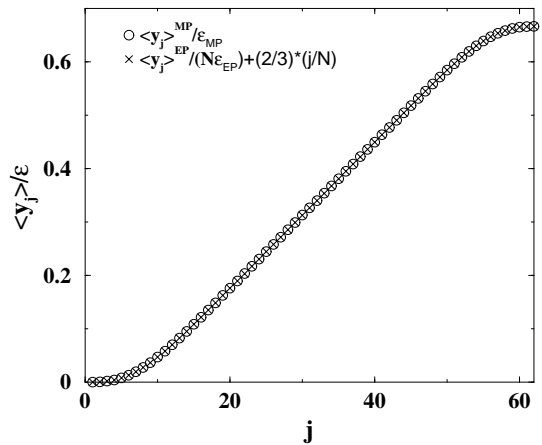}
 \caption{Comparison of the MP orientation with a translated EP values for 
the same value $N=63$ and $d=1$}  \label{compar}
\end{figure}
Although the MP orientation is our basic ingredient, plotting the EP orientations more clearly
shows the $N$ dependence, see Fig. \ref{rawrd}. To explain the complicated scaling behavior 
of the these curves is the main aim of this paper. We already note  that 
the curves (almost) pass through the same point at the head and tail, since there is a well known
relation \cite{Kooiman,vanLeeuwen} between the orientation of the first link and the drift velocity.
\begin{equation} \label{b3}
\langle y_1 \rangle = -2 \epsilon /3 + v_{EP} = -2\epsilon/3 + \epsilon/3N. 
\end{equation} 
\begin{figure}[h]
\centering
\includegraphics[width=7cm]{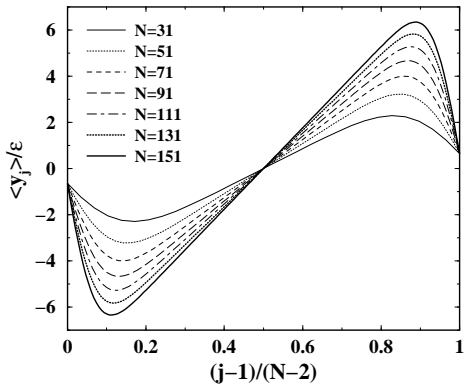}
\caption{Orientations for EP chains of length $N=31$ to 151 at  $\epsilon = 0.00005$ and $d=1$} \label{rawrd}
\end{figure}
 
In first order, the field dependence 
only combines with the sign of one link at the time
\begin{equation} \label{b4}
P^1_{MP} ({\bf y}) = \sum^{N-1}_{j=1}  y_j \, P_j ({\bf x}).
\end{equation} 
Here ${\bf x}$ is the vector $(x_1, \cdots , x_{N-1})$ with $x_j = y^2_j$. The
remaining factor $P_j ({\bf x})$ depends on the position $j$ of this link but not on the
signs of the other links. To distinguish link $j$ from the others we give it a tag.  
The notation $P_j ({\bf x})$ anticipates that it can be seen as the (unnormalized) 
probability that the chain is in configuration ${\bf x}$ and the tagged link at $j$.
Relation (\ref{b4}) reduces the relevant configuration space from the original $3^{N-1}$
points to $(N-1)2^{N-1}$ points $(j,{\bf x})$.
We refer to $x_i =1$ as a ``particle'' and to $x_i = 0$ as a ``hole''. The marked link $j$
is a tagged particle. 

The probability interpretation for $P_j ({\bf x})$ is unique for the MP model. 
As (\ref{b1}) shows, all $P^1$ can be cast in the form (\ref{b4}), 
but the corresponding $P_j$'s do not need to be positive. 
The equation for $P_j ({\bf x})$ follows from (\ref{a3}), summing over
all ${\bf y}$ which lead to the same configuration $(j, {\bf x})$. The known term in (\ref{a3}) 
becomes with $L = \sum_j x_j$,
\begin{equation} \label{b5}
{\cal M}^1_{MP} P^0 ({\bf y}) = y_j \delta_{j, N-1} \, P^0 ({\bf x}), \quad 
P^0 ({\bf x}) = 2^L/ 3^{N-1},
\end{equation}
as only the last repton is effected by the field. We can eliminate this term from (\ref{a3})
using a symmetry of the EP model. When all reptons are pulled evenly, the transformation
\begin{equation} \label{b6}
{\bf y} = (y_1, \cdots , y_{N-1}) \leftrightarrow {\bf y}^T =   (- y_{N-1}, \cdots , - y_1), 
\end{equation}
leaves the probability invariant, as the numbering from head to tail is equivalent to our
numbering from tail to head. This symmetry is clearly reflected in the curves of Fig. \ref{rawrd}.
With relation (\ref{b1}) we can transfer the EP symmetry to the MP model. It yields
\begin{equation} \label{b7}
P_j ({\bf x}) + P_{N-j} ({\bf x}^T) = P^0 ({\bf x}).
\end{equation}
The transposed vector is ${\bf x}^T = (x_{N-1}, \cdots , x_1)$. We use (\ref{b7}) for $j=N-1$,
in order to eliminate the known term in (\ref{a3}). 
Then equation (\ref{a3}) for $P_j ({\bf x})$ becomes
homogeneous again and therefore a stationary Master Equation for  $P_j ({\bf x})$, which
proves that $P_j ({\bf x})$ can be interpreted as a probability. This new Master Equation is
not very different from the original one (\ref{a1}). The substitution has added an extra process:
the tagged particle may jump from the tail to the head, thereby changing the configuration
$(1, {\bf x})$ to $(N-1, {\bf x}^T)$. In addition we have the rule that the tagged particle
may not leave or enter the chain, as a consequence of the fact that the tagged particle
always has to be present.

This  interpretation describes $P_j ({\bf x})$ as a random walk of a 
tagged particle at $j$ in the sea of other particles and holes in configuration ${\bf x}$.
The possible moves are interchanges of particles and holes. 
Particles and holes may leave and enter the chain, but not the tag,
which can jump from tail to head (and not reverse) as described above. The probability 
$p_j$ that the tag is at $j$, is $\sum_{\bf x} P_j ({\bf x})$. 
This is the quantity to be discussed, since it is equivalent to the local orientation:
$\langle y_j \rangle = 2 p_j /3$, the 2/3 coming from the fraction of particles in the chain. 

An equation for $p_j$ follows from the observation that the tag can only jump to neighboring 
positions $j \pm 1$, leading to the effective equation
\begin{equation} \label{c1}
W_+ (j-1) \, p _{j-1}+ W_- (j+1)\,  p_{j+1} = [ W_+ (j)+W_- (j)]\,  p_j, 
\end{equation} 
with appropriate changes at the end links. $W_+ (j)$ is the probability that, 
if the tag is at $j$, a hole is at $j+1$ such that the tag can jump to the right. 
Similarly $W_- (j)$ is the conditional probability that the tag at $j$ is neighbored 
by a hole at $j-1$. Fortunately the DMRG calculations
not only provide information on $p_j$, but also on the transition rates
$W_\pm (j)$, since $W_\pm (j) p_j$ is the joint probability on a tag-hole pair and these 
correlation functions have been calculated \cite{Carlon2}. 
If the holes were randomly distributed, one would expect 1/3 for these rates, a value to which 
the rates converge in the bulk of the chain. The deviations from 1/3 are the largest at the 
tail of the chain (not at the head!). After a short initial regime they start to fall off 
smoothly as $1/j$. In the smooth regime one may replace the difference equation (\ref{c1}), 
by a Fokker--Planck equation, keeping only the first and second derivative with respect to $j$.

The essential point in the scaling analysis is that the difference of the $W$'s obeys a 
simple scaling relation
\begin{equation} \label{c2}
[W_+ (j) - W_- (j)] \simeq g(x) / j, \quad \quad \quad x=j / \sqrt{N},
\end{equation}
which is demonstrated in Fig. \ref{scaling}, where we plot the difference against the scaling
variable $x=j / \sqrt{N}$.
\begin{figure}[h]
\centering   
\includegraphics[width=7cm]{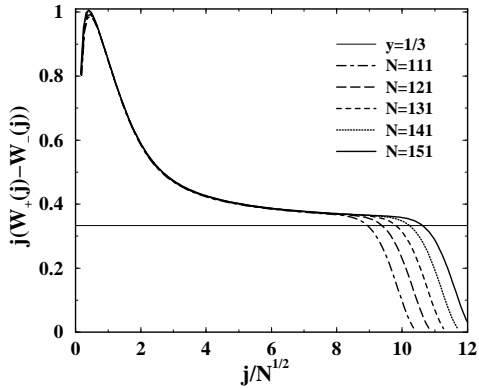}
  \caption{The scaling form for the difference (\ref{c2})}  \label{scaling}
\end{figure}
This difference enters in the Fokker-Planck equation as the systematic force. It tends to
push the tag away from the tail.
For $N \rightarrow \infty$, the part at the head, where the various values of $N$ 
fan out, shifts to larger and larger values, while the function $g(x)$ approaches the limit 1/3. 
The small region at the tail side, where the scaling does not hold, shrinks with increasing $N$ 
in this plot. In this region the transition rates change appreciably with each link, and one
has to use the discrete equations (\ref{c1}), starting from a value $p_1$. 

For values $j \sim \sqrt{N}$ the Fokker-Planck equation reads, writing $p_j = p(j/\sqrt{N})$
\begin{equation} \label{c3}
\frac{d}{dx} \left[\frac{g (x)} {x} p(x)\right] = \frac{1}{3} \frac{d^2 \, p (x)}{dx^2 }. 
\end{equation} 
In the right hand side we have replaced the sum of the $W$'s by 2/3, since the correction
is an order $\sqrt{N}$ smaller. The dominant solution of (\ref{c3}) has the form
\begin{equation} \label{c4}
p(x) = p(x_0) e^{R(x)}, \quad R (x) = 3 \int^x_{x_0} dx' \,g(x')/x'.
\end{equation} 
$x_0$ is the point where we match the solution with the outcome of the intial regime. 

The solution (\ref{c4}) bridges the initial and asymptotic behavior.
We first inspect what (\ref{c4}) gives for the middle of the chain 
$x_m = \sqrt{N} /2$. From the asymptotics of $g(x)$ for small and large $x$ one finds
\begin{equation} \label{d1}
p (x_m) \sim p (x_0) \, N^{3 [g(0) + g(\infty)]}.
\end{equation}  
Now $p (x_m) = 1/2$ by symmetry and $p (x_0)$ inherits its magnitude from $p_1$, which
equals $3 v_{MP}/2$ and is therefore order $N^{-2}$ (see (\ref{b3}) and 
\cite{vanLeeuwen,Spohn}). Thus we find for the exponent
\begin{equation} \label{d2} 
3 \, [ g(0) + g(\infty)] = 4.
\end{equation} 
The value of $g(\infty)$ follows from the behavior of $R(x)$ around $x_m$. Expansion of
$g(x)$ in powers of $x^{-1}$ gives
\begin{equation} \label{d4}
p_j \simeq \frac{1}{2} \left( \frac{2j}{N}\right)^{3 g(\infty)} \left[1 +3g_{-1} \left( \frac{2}{\sqrt{N}} -
\frac{\sqrt{N}}{j} \right) \right] 
\end{equation}
Thus $g(\infty) = 1/3$, otherwise the behavior would not be linear in $j/N$. With
(\ref{d2}) this implies $g(0) =1$.
The $1/\sqrt{N}$ terms lead to the slope of the EP profile in the middle. 
This slope must grow as $\sqrt{N}$ and a numerical fit gives $g_{-1} = 0.5$.
The small $x$ behavior of $R(x)$ yields
\begin{equation} \label{d5}
p(x) \simeq p(x_0)\left( x /x_0 \right)^3.
\end{equation} 
This power is higher than the estimate of Barkema and Newman\cite{Barkema2}, who 
give  2.7 for the exponent. We could not univocally find an exponent by fitting the initial
orientation of a long chain.
Here we see that the power 3 is fixed by the asymptotic properties of the scaling function $g(x)$, 
for small and large $x$, following from $v_{MP}$ and the slope
in the middle of the chain. We also note that this power does not set in inmediately at the tail,
but develops in the early regime of the variable $x=j/\sqrt{N}$. Even for our longest chains 
$N=151$, the window where the power applies is rather small.

In summary, the regions of order $\sqrt{N}$ at the ends of the chain which differ
from the bulk, originate from the algebraic decay (as $1/j$) towards the bulk of the 
correlations between two successive links. 
This explains why finite size corrections decay with an exponent $-1/2$. 

{\bf Acknowledgments}.
This work has been supported by the Polish Science Committee (KBN) 
through Grant No. 2 P03B 125 24. JMJvanL wants to thank Marcel den Nijs for extensive
and stimulating discussions.

\end{document}